\documentclass[prl,twocolumn,aps,a4paper,superscriptaddress,showpacs]{revtex4}
\usepackage{amsmath,latexsym,amsfonts}
\usepackage[dvips]{graphicx}
\begin{document}
\title{Can be gravitational waves markers for an extra-dimension?}

\author{Emanuele Alesci}
\email{emanuelealesci@icra.it}
\author{Giovanni Montani}
\email{montani@icra.it} 
\affiliation{Dipartimento di Fisica Universit\`a di Roma ``La Sapienza''}
\affiliation{ICRA---International Center for Relativistic Astrophysics  
c/o Dipartimento di Fisica (G9) Universit\`a di Roma ``La Sapienza'',
Piazza A.Moro 5 00185 Rome, Italy}

\today

\begin{abstract}
The main issue of the present letter is to fix specific features (which turn out being independent of extradimension size) of gravitational waves generated before a dimensional compactification process. Valuable is the possibility to detect our prediction from gravitational wave experiment without high energy laboratory investigation. In particular we show how gravitational waves can bring information on the number of Universe dimensions.
Within the framework of Kaluza-Klein hypotheses, a different morphology arises between waves generated before than the compactification process settled down and ordinary 4-dimensional waves. In the former case the scalar and tensor degrees of freedom can not be resolved. As a consequence if were detected gravitational waves having the feature here predicted (anomalous polarization amplitudes), then they would be reliable markers for the existence of an extra dimension. 

\end{abstract}



\maketitle

\emph{I. Introduction.}-
In recent years wide interest renewed on the multidimensional nature of the Universe, essentially because of the intense investigation on brane theories \cite{Maartens1}. Thus a great effort has been done in order to outline same phenomenological issue for extradimension \cite{Optical, EvolutCosmologicalperturbation, Transdim phys in inflation, Sub-Lunar PBH, gravitoncosmologyinUED}. Our work stand in this latter line of research and face the study of waves propagation on a 5d universe (for connected work see \cite{EvolutCosmologicalperturbation, Transdim phys in inflation, versusgraviton, gravitonproductioned, Generalized waves in any dim, Gravitational Radiation in D-dim, Detecting extra dimensions with gravity wave spectroscopy}) as well as to fix features detectable from gravitational waves experiment like the actual LISA project \cite{Lisa}. In fact the observation of an extradimension requires so large energy that only cosmological phenomena are suitable for its observation; using gravitational waves as markers allows,(in view of their very early decoupling from the primordial equilibrium) to detect phenomena of huge energy scale and so to test pre-compactification process. Adding a spacelike dimension to the space-time allows to geometrize the electromagnetic field via the extradimensional degrees of freedom associated to the 5-dimensional metric tensor \cite{1, 2}. This relevant issue is achieved paying the price to restrict either the structure of the space-time (to be taken as the direct sum of a generic 4-dimensional manifold and a circle compactified to very small lengths (see \cite{stimadimensioni} for current experiment limit), either on the admissible coordinates transformations (only translations have to take place along the extra-dimension). Thus in the framework of a Kaluza-Klein (KK) theory \cite{MKKT, Witten, Wesson-rep}, to include the electromagnetic field into the spacetime geometry requires that the 5-dimensional Principle of General Relativity is explicitly broken down. The appearance of a dimensional compactification process can be explained by anisotropic Universe dynamics \cite{gone?}, but the most natural way to restore the Principle of General Relativity within a KK approach, consists of involving the so-called ``Spontaneous Compactification mechanism''. According to this idea \cite{Sch, Crem1, Crem2, Luciani}, the 5-dimensional theory is yet governed by the Einstein-Hilbert action, but the full Poincar\'e invariance is spontaneously broken because the ``vacuum state'' has the structure of a 4-dimensional Minkowsky space plus a compactified circle.	Thus we may argue that during the Universe evolution, existed a stage (whose temperature is expectable between $10^{15}GeV$ and $10^{19}GeV$) in which the spontaneous symmetry breaking of the Poincar\'e group took place and the Universe settled down into its vacuum state, i.e. an expanding homogeneous and isotropic background plus a compactified circle. 
In the present work we analyze the different behavior existing between a gravitational perturbation which is generated before the process of spontaneous compactification has taken place and an ordinary gravitational wave \cite{MTW}. By other words we investigate if observing  4-dimensional gravitational waves, it is possible to recognize features which are a consequence of the extra-dimension. 
To this end we consider a 5-dimensional gravitational wave (with its field equations and gauge conditions) and constraint it by KK restrictions. The main issue of our investigation is that in this context, the electromagnetic wave (spin 1 component) has the usual structure, while the 4-dimensional gravitational waves (spin 2 component) and the scalar perturbation (spin 0 component) correlate.
When the gravitational wave is generated in five dimension, before compactification, there is no longer possibility to split its spin 2 and spin 0 components into independent degrees of freedom. As a consequence the 4-dimensional gravitational waves have different amplitude of oscillation along the two polarization states. Such a feature is an effect of the mixing between the scalar and tensor degrees of freedom and therefore is a reliable marker for the existence of a KK configuration of the Universe. In fact we stress that 4-waves can always be taken in Transverse-Traceless gauge and therefore the anomalous behavior we will find here (waves are no longer traceless) can not arise from the natural coupling with ordinary matter fields.
We underline that the result here presented is not affected by the size of the extradimension.

\emph {II. 5-d wave on a KK background}-
We consider a gravitational wave propagating through a 5-dimensional vacuum background. We use a perturbative approach which allows us to split the metric into background coefficients $j_{AB}$ and perturbation terms $h_{AB}$:
\begin{equation}
	^{5d}g_{AB}=j_{AB}+h_{AB} \qquad(A,B=0,1,2,3,5) \label{1}
\end{equation}
The Ricci tensor for a metric of the form \eqref{1} can be splitted into
$R_{AB}=R^{(0)}_{AB}+R^{(1)}_{AB}(h)+O(h^2)$ 
where $R^{(0)}_{AB}$ is built with the background metric $j_{AB}$ and $R^{(1)}_{AB}$ is the first order correction in $h_{AB}$. 
We are looking for a linearized theory (a wave that doesn't affect its own propagation) therefore we neglect the second order terms.
In the vacuum $R_{AB}=0$ and, since the metric $j_{AB}$ is a vacuum background, then $R^{(1)}_{AB}=0$ (hereafter, the indices are raised and lowered with the unperturbed metric $j_{AB}$):
\begin{equation}
R^{(1)}_{AB}=\frac{1}{2}(h^C_{\;A;B;C}	+h^C_{\;B;A;C}-h_{AB\phantom{C};C}^{\phantom{AB};C}-h_{;A;B})=0\label{4}
\end{equation}
This is the propagation wave equation for a 5-dimensional gravitational wave (the covariant derivatives refers to the background metric $j_{AB}$).\\
Introducing the tensor $\psi_{AB}=h_{AB}-\frac{1}{2}j_{AB}h$ ($h=h_{AB}j^{AB}$) 
the equation \eqref{4} becomes
\begin{equation}
\begin{split}
&-\psi^{\phantom{AB};C}_{AB\phantom{\;C};C}-j_{AB}\psi^{AB}_{\phantom{AB};A;B}+2\psi^C_{\;(A;C;B)}-2 R^{(0)}_{CADB}\psi^{CD}=0\label{5}
\end{split}
\end{equation}
where $_{()}$ is the symmetric tensor and $R_{CADB}$ is the 5-d Riemann tensor built with $j_{AB}$. 
With a particular choice of gauge we can simplify the last expression: we can make an infinitesimal coordinate transformation $x'^A=x^A+\xi^{A}$ that induces a first order change on the perturbation
\begin{equation}
	h_{AB}\longrightarrow h'_{AB}=h_{AB}-\xi_{A;B}-\xi_{B;A}\label{6}
	\end{equation}
This gauge freedom can be used to impose the "Hilbert gauge" $\psi^{\;\;B}_{A\;\,;B}=0\label{7}$.
In this gauge the equation \eqref{5} becomes:
\begin{equation}
 \left\{
\begin{aligned}
	&-\psi^{\phantom{AB};C}_{AB\phantom{\;C};C}-2 R^{(0)}_{CADB}\psi^{CD}=0\\
 &	\psi^{\;\;B}_{A\;\,;B}=0
\end{aligned}\right.\label{8}
\end{equation} 
We have fixed our attention on a 5-dimensional gravitational wave because, if the wave was generated in a 5-d universe, before a spontaneous compactification process had taken place, then it would be a purely  5-dimensional object, generally covariant under arbitrary 5-d coordinate transformations.

After the spontaneous compactification takes place, the Universe acquires a KK structure (the manifold $M^{5}$ becomes $M^{4}\times S^1$) and the 5-d local Poincar\'e group is spontaneously broken into a 4-d local Poincar\'e group and a U(1) local gauge group.
The wave, originally 5-dimensional, now feels the effects of the compactification and its components transform in a different way under 4-d coordinate transformations. 
In unperturbed KK theory indeed, the metric tensor $j_{AB}$ has the following decomposition \cite{MKKT,Wesson-rep}:
\begin{equation}
\begin{split}
	j_{55}=\Phi^2\qquad j_{5\mu}=ek\Phi^2A_\mu \\
	j_{\mu\nu}={}^{4d}g_{\mu\nu}+e^2k^2 \Phi^2A_\mu A_\nu
\end{split}\label{9}
\end{equation}
where $\Phi^2$ is a scalar function, $A_{\mu}$ is the electromagnetic field and ${}^{4d}g_{\mu\nu}$ is the gravitational field, $e$ is the electric charge and $k$ is a dimensional constant. In this theory all the fields $\Phi^2$, $A_{\mu}$, $^{4d}g_{\mu\nu}$ are purely 4-d objects and are independent of the extra-dimension coordinate $x^5$.
In the same way the following identifications will be correct for the whole components of ${}^{5d}g_{AB}$ in a perturbed theory:
\begin{equation}
\begin{split}
		{}^{5d}g_{55}=\Phi^2+h_{\Phi} \quad \frac{{}^{5d}g_{5\mu}}{{}^{5d}g_{55}}=A_\mu+\epsilon_{\mu}\\
		{}^{5d}g_{\mu\nu}-\frac{{}^{5d}g_{5\mu}{}^{5d}g_{5\nu}}{{}^{5d}g_{55}}={}^{4d}g_{\mu\nu}+\epsilon_{\mu\nu}
\end{split}
\label{11}
\end{equation}
where the infinitesimal fields $h_{\Phi}$, $\epsilon_{\mu}$, $\epsilon_{\mu\nu}$ are respectively a 4-d scalar field, a 4-d vector field and a 4-d tensor field.
 
To now understand how the original 5-d wave splits itself in 4-d objects after the compactification, we must look at the propagation equation \eqref{8} and extract from it the purely 4-d quantities. In order to do this we must extract all the 4-d geometrical objects (Christoffel, Riemann, Ricci) contained in 5-d geometrical objects and extract the 4-d fields, contained in the 5-d field $\psi_{AB}$, because, after the spontaneous compactification, the 5-d general covariance is lost, and the components of $\psi_{AB}$ acquire a different behavior under 4-d general coordinate transformations (becoming distinct 4-d dynamical fields).  
We will base our analysis on the following fixed background $j_{AB}$:
\begin{equation}
	j_{AB}=\left( \begin{array}{cc}
	 {}^{4d}g_{\mu\nu}  & 0 \\
	0 & 1 \\
	\end{array}\right)
	\label{12}
\end{equation}
where we have chosen $\Phi^2=1$ in the spirit of the Kaluza approach (see \cite{MKKT,Wesson-rep} for a detailed discussion on the term $\Phi^2$ ) and $A_{\mu}=0$ because we are looking for a wave that propagates in a cosmological background in which a large scale electromagnetic field is absent. 
Calculating the 5-d geometrical objects  and remembering the cylindricity condition ($\partial_{5}=0$), is easy to check that all 5-d Christoffel and 5-d Riemann components with 5 index are null and the components with index 4-dimensional coincide with the 4-d object (built with ${}^{4d}g_{\mu\nu}$).
The last step is to extract the 4-d dynamical fields contained in the components of $\psi_{AB}$; using the equations \eqref{11} and substituting the metric \eqref{12} in the total metric ${}^{5d}g_{AB}$ we obtain up to the first order the following identifications: $h_{55}=h_{\Phi}$, $h_{5\mu}=\epsilon_{\mu}$, $h_{\mu\nu}=\epsilon_{\mu\nu}$.
By the use of such identifications we can finally decompose the tensor $\psi_{AB}$ in four dimensional objects:
 \begin{equation}
\left\{\begin{aligned}
&\psi_{55}=h_{55}-\frac{1}{2}j_{55}h=h_{\Phi}-\frac{1}{2}(h_{\Phi}+\epsilon)= \frac{h_{\Phi}-\epsilon}{2}\\
&\psi_{5\mu}=h_{5\mu}-\frac{1}{2}j_{5\mu}h=\epsilon_{\mu}\\
&\psi_{\mu\nu}=h_{\mu\nu}-\frac{1}{2}j_{\mu\nu}h=\epsilon_{\mu\nu}-\frac{1}{2}{}^{4d}g_{\mu\nu}(h_{\Phi}+\epsilon)
	\end{aligned}\right.\label{19}	
\end{equation}\\    
where $\epsilon= {}^{4d}g_{\mu\nu}\;\epsilon^{\mu\nu}$ and $h=j_{AB}\;h^{AB}=h_{\Phi}+\epsilon$.
Substituting the 5-d components of $\psi_{AB}$ with the \eqref{19} inside the propagation equations for 5-d gravitational wave \eqref{8} and in the line of what we have said about 5-d Riemann and Christoffel we obtain 
\begin{equation}
\left\{\begin{aligned}
&\left({h_{\Phi}-\epsilon}\right)^{;\mu}_{\phantom{;\mu};\mu}=0\\
	&-\epsilon^{\phantom{\mu};\nu}_{\mu\phantom{;\nu};\nu}=0 \\
&-\epsilon^{\phantom{\mu\nu};\rho}_{\mu\nu\phantom{\rho};\rho}+\frac{1}{2}{}^{4d}g_{\mu\nu}(h_{\Phi}+\epsilon)^{;\rho}_{\phantom{;\rho};\rho}+\\
	&-2\;\, {}^{4d}R_{\rho\mu\sigma\nu}\epsilon^{\rho\sigma}+{}^{4d}R_{\mu\nu}\;(h_{\Phi}+\epsilon)=0 
\end{aligned}\right.
\label {20}
\end {equation}
and their gauge equations:
\begin{equation}
\left\{\begin{aligned}
	&\epsilon^{\rho}_{\phantom{\rho};\rho}=0  \\
&\epsilon_{\mu\phantom{\rho};\rho}^{\phantom{\mu}\rho}-\frac{1}{2}\epsilon_{;\mu}=\frac{1}{2}h_{\Phi;\mu}  
\end{aligned}\right.	         \label {21}
\end{equation}
To simplify these expressions we must remember that the original 5-d wave was propagating in the vacuum (${}^{5d}R_{\mu\nu}={}^{4d}R_{\mu\nu}=0$) and if we contract the last of the \eqref{20} with the unperturbed 4-d metric ${}^{4d}g_{\mu\nu}$ we obtain $h^{\phantom{\Phi};\rho}_{\Phi\phantom{;\rho};\rho}=-\frac{1}{2}\epsilon^{;\rho}_{\phantom{;\rho};\rho}$. Using these considerations the propagation system \eqref{20} for the 4-d fields $h_{\Phi}$, $\epsilon_{\mu}$, $\epsilon_{\mu\nu}$
becomes 
\begin{equation}
	\left\{\begin{aligned}
&h^{\phantom{\Phi};\rho}_{\Phi\phantom{;\rho};\rho}=\epsilon^{;\rho}_{\phantom{;\rho};\rho}=0\\
	&-\epsilon^{\phantom{\mu};\nu}_{\mu\phantom{;\nu};\nu}=0 \\
&-\epsilon^{\phantom{\mu\nu};\rho}_{\mu\nu\phantom{\rho};\rho}-2\;\,{}^{4d}R_{\rho\mu\sigma\nu}\epsilon^{\rho\sigma}=0 
\end{aligned}\right.
\label {24}
\end{equation}
The equations \eqref{24} and \eqref{21} are the wave's equations (the covariant derivatives are four dimensional) on a 4-d vacuum background ${}^{4d}g_{\mu\nu}$ of three different fields with fixed gauge, a massless scalar field $h_{\Phi}$, a massless vector field $\epsilon_{\mu}$ in Lorentz gauge $\epsilon^{\rho}_{\phantom{\rho};\rho}=0$, and a massless tensor field $\epsilon_{\mu\nu}$ in a ``new'' gauge resembling the Hilbert gauge but having a coupling with the scalar wave.
 
We conclude that a 5-d gravitational wave, after the compactification, can be seen as a superposition of a scalar, a vector and a tensor wave. The scalar wave does not have a direct physic interpretation but the vector and the tensor wave can be identified with a 4-d electromagnetic wave and a 4-d gravitational wave. To proceed with this identification we must verify that the gauge freedom of the 5-d field $h_{AB}$ becomes the right gauge freedom for these 4-d fields. 
The infinitesimal coordinate change $x'^A=x^A+\xi^A$ generates the transformation \eqref{6} on the perturbation $h_{AB}$. To understand how this gauge freedom operates on the 4-d fields, we must analyze the 5d-vector $\xi^A$; when the 5-d general covariance is lost, the admissible coordinate change restricts to  
\begin{equation}
x^5=x'^5+f(x'^\nu)\qquad \qquad	
x^\mu=x^\mu(x'^\nu)
\label{27}
\end{equation}
and the transformation $x'^A=x^A+\xi^A$ must be of the same kind \eqref{27} too; this implies that the components of $\xi^{A}$ with indices $\mu$ must be a 4-vector and that $\xi^5$ must be a scalar function.
Using this decomposition of the vector $\xi^A$ and remembering that, $\nabla_{5}=0$,${}^{5d}\nabla_\mu={}^{4d}\nabla_\mu$ the components of the transformation in \eqref{6} become 
\begin{align}
&h_{55}\longrightarrow h'_{55}=h_{55}\label{28}\\
	&h_{5\mu}\longrightarrow h'_{5\mu}=h_{5\mu}-\xi_{5,\mu}\label{29}\\
&	h_{\mu\nu}\longrightarrow h'_{\mu\nu}=h_{\mu\nu}-\xi_{\mu;\nu}-\xi_{\nu;\mu}\label{30}
\end{align}
Using the identifications $h_{55}=h_{\Phi}$, $h_{5\mu}=\epsilon_{\mu}$, $h_{\mu\nu}=\epsilon_{\mu\nu}$, we can say that the original 5-d gauge freedom splits into \eqref{28} which shows the absence of a gauge freedom for $h_{\Phi}$, confirming its scalar nature, \eqref{29} which shows (being $\xi^5$ a scalar function) for $\epsilon_{\mu}$ the same gauge freedom of an electromagnetic field, \eqref{30} which shows for $\epsilon_{\mu\nu}$ ($\xi^\mu$ is a 4-vector and the covariant derivatives are now built with ${}^{4d}g_{\mu\nu}$) the same gauge freedom of an ordinary 4-d gravitational wave.
We know that $h_{\Phi}$, $\epsilon_{\mu}$, $\epsilon_{\mu\nu}$ have wave  equations, of scalar, electromagnetic and gravitational fields  respectively and that they also have the right behavior under gauge transformations.
Looking at the degrees of freedom we can confirm our identifications.
In the pre-compactification framework the field $\psi_{AB}$ has 15 components; the gauge $\psi^{\;\;B}_{A\;\,;B}=0$ leaves only 10, but we still have the freedom of make an other transformation \eqref{6}, such as $\xi_{A\phantom{B};B}^{\phantom{A};B}=0$ that leaves only 5 independent components for $\psi_{AB}$.
We have taken as a starting point for the wave, before compactification, the system \eqref{8} which leaves the wave with 10 degrees of freedom. After compactification, when the general covariance is lost, the field $h_{AB}$ splits its degrees of freedom between 4-d fields; the second gauge transformations, whose generators satisfy $\xi_{\phantom{\mu};\mu}^{5\phantom{\mu};\mu}=0$ and $\xi_{\phantom{\mu};\mu}^{\nu\phantom{\mu};\mu}=0$, allows us to make further transformations that ensure one degree of freedom for  $h_{\Phi}$, two for $\epsilon_{\mu}$ and two for $\epsilon_{\mu\nu}$ making consistent the identification with a scalar wave, an electromagnetic wave and a gravitational wave.      

\begin{figure}
\includegraphics[width=3.8cm,]{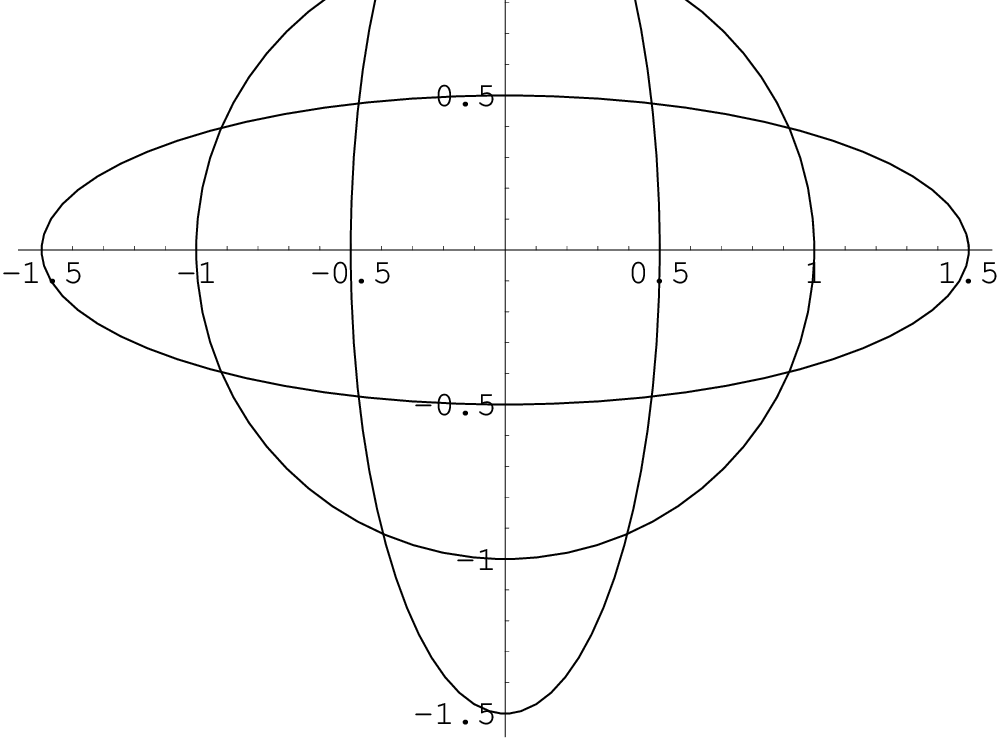}
\includegraphics[width=3.8cm,]{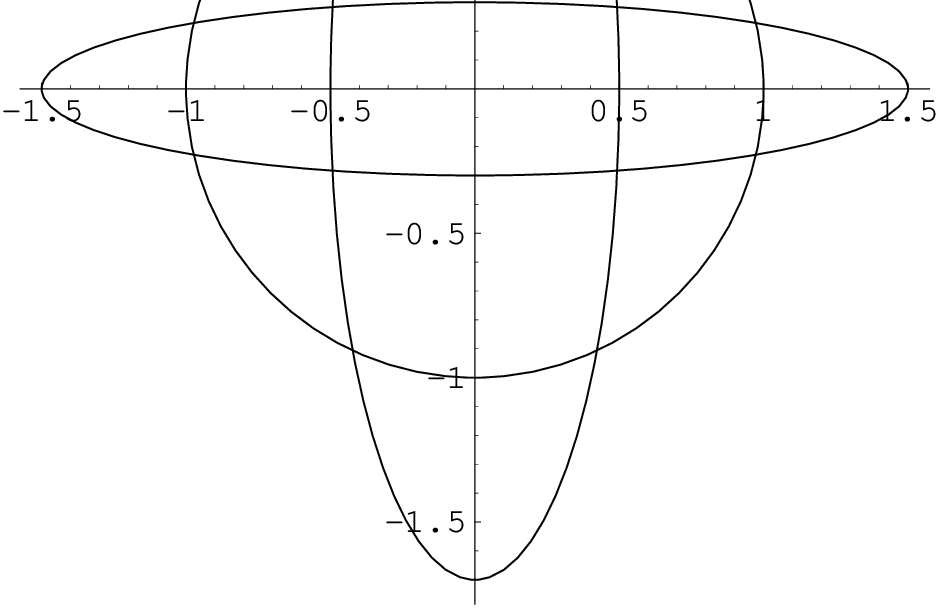}
\caption{Deformations of a test particle ring produced by gravitational waves: the first image is for an usual gravitational wave, the second one is for the wave $\epsilon_{\mu\nu}$ with 5-d origin (in the figure the field $h_{\Phi}$ has an amplitude of 0.2 and $C_{11}$ is 0.5)}
\label{fig1}
\end{figure}

\emph{III. 4-d gravitational wave with 5-d origin}- We now show how the ``strange'' gauge (the second one in \eqref{21}) affects the morphology of the 4-d gravitational wave $\epsilon_{\mu\nu}$ with 5-d origin. We study the particular case of the flat 5-d metric $j_{AB}=\eta_{AB}$ (having in mind that gravitational waves are actually detected in Minkowsky space-time).
We can take as solution of the system \eqref{24} the plane waves $\epsilon_{\mu\nu}=\Re e \left\{C_{\mu\nu} \;\;e^{ik_\alpha x^\alpha}\right\}$,
		$\epsilon_{\mu}=\Re e \left\{C_{\mu} \;\;e^{ik_\alpha x^\alpha}\right\}$ and 		$h_{\Phi}=\Re e \left\{\phi \;\;e^{ik_\alpha x^\alpha}\right\}$
which must satisfy the following conditions imposed by \eqref{21}:
\begin{equation}
\begin{split}
	k_\mu k^\mu=0\qquad
	C_\mu k^\mu=0\\
	k^\mu C_{\mu\nu}-\frac{1}{2}Ck_{\nu}=\frac{1}{2}k_\nu\phi 
\end{split}
\label{39}
\end {equation}
where $\phi$, $C_\mu$ and $C_{\mu\nu}$ are taken as constants   
(we have chosen a solution with the only wave's vector $k^\mu$ to develop the idea of a single 5-d wave that splits its component cause the KK structure). The electromagnetic wave $\epsilon^{\mu}$ in Lorentz gauge is independent by the gravitational wave and we can restrict our analysis to the third expression in \eqref{39}.     
If we take a wave that propagates in $\hat{3}$ direction the equations \eqref{39} allow to express the components $C_{0i}$ and $C_{22}$ as function of the other components 
\begin{equation}
\begin{split}
	&C_{01}=-C_{31}\qquad \qquad \quad C_{02}=-C_{32}\\
	&C_{03}=-\frac{1}{2}(C_{33}+C_{00})\quad
	C_{22}=-C_{11}-\phi
\end{split}
\label{41}\end{equation}
We can still induce the transformation  $\epsilon_{\mu\nu}\rightarrow \epsilon'_{\mu\nu}=\Re e \left\{C'_{\mu\nu}\;e^{ik_\alpha x^\alpha}\right\}$ using as generator $\xi^\mu(x)$ the infinitesimal 4-d vector (which satisfies $\Box\xi^\mu=0$) $\xi^\mu(x)=i\chi^\mu e^{ik_\alpha x^\alpha}-i\chi^{\mu}e^{-ik_\alpha x^\alpha}$
where $\chi^\mu$ is constant.
Choosing the components of the vector $\chi^\mu$ we can cancel the components $C_{3i}$, $C_{00}$ and, as a consequence of the \eqref{41}, the $C_{0i}$ too. 
Exhausted the gauge freedom the polarization tensor, for the presence of the field $h_{\Phi}$, is
  \begin{equation}
	C'_{\mu\nu}=\left( \begin{array}{cccc}
	0 & 0 & 0 & 0 \\
	0 &  C_{11} &  C_{12}&0 \\
	0 & C_{12}& -C_{11}-\phi& 0 \\ 
	0 & 0 &0 & 0\end{array}\right) 
 \label{44}
 \end{equation}  
We can note that, in spite of we have performed the procedure leading to the TT-gauge, the ``strange'' gauge condition prevent to eliminate the trace $C'\equiv C'_{\mu\nu}\eta^{\mu\nu}=-\phi$. 
We can now study the testable effects of this anomalous gravitational wave looking at the geodesic deviation.   
If we consider a particle A at rest in the origin of the coordinate system with the  $\hat {3}$ axis in the direction of propagation of the incoming wave and chosen to put the polarization tensor in the form \eqref{44}, and a particle B disposed at a distance $\delta x^\mu$ from the particle A, the geodesic deviation between the two particles will be (neglecting second order terms)
\begin{equation}
	\frac{D^2\delta x^\mu}{d\tau^2}=\frac{1}{2}\eta^{\mu i} \epsilon_{ij,0,0} \; \delta x^j \qquad\textrm{with ij=1,2}\label{45}
\end{equation}   
If we consider the case $\epsilon_{12}\neq0$ and $\epsilon_{11}=\epsilon_{22}=0$ this equation is the same for an ordinary gravitational wave, but if we consider the opposite case $\epsilon_{12}=0$, $\epsilon_{11}\neq0$ and $\epsilon_{22}\neq0$, in spite of the geodesic deviation equation is the same of the usual 4-d theory, the component $\epsilon_{22}\neq-\epsilon_{11}$. This fact implies that, imagining of dispose a test particles ring, the passage of the wave will cause an oscillation different from the usual one: while the ring, under the effect of an usual wave, would oscillate each period quarter of the incoming wave into an ellipse, a circle, then into the same ellipse rotated by $90^{\circ}$ and so on; in the present case the ring would oscillate in deformed ellipses with a different axes elongation (they aren't the same ellipse rotated by $90^{\circ}$)(in fig. \ref{fig1} is shown the comparison between an ordinary gravitational wave and the wave with 5-d origin).

\emph{IV. Concluding remarks}- The merit of the analysis above presented relies on outline strong qualitative features which have to characterize a space-time perturbation generated before the Universe underwent a compactification process. These features are independent on the size of the extradimension and they expectably extend to any multidimensional space-time. But, overall, we showed how a phenomenology for an extradimension can be predicted by observing gravitational waves without requiring high energy experiment via accelerators; this fact is allowed because gravitational wave can arise from the very early and high temperature Universe and then freely propagate to us because of their weak coupling. Specifically the impossibility to separate the spin 0 and spin 2 components of the dimensional reduced perturbation was predicted and phenomenological implications were discussed:       
the 5-d origin produces effects on gravitational waves that can not arise from ordinary 4-d space-time geometry. In particular pre-compactification waves exhibit anomalous polarization states; such gravitational waves, once detected, could provide an indication for the existence of an extradimension.


\begin{thebibliography}{}
\bibitem{Maartens1}
R. Maartens,  \emph{Brane-world gravity}, Living Reviews Relativity 7 (2004) gr-qc/0312059

\bibitem{Optical}
A. K. Ganguly, R. Parthasarathy, Phys. Rev. D {\bf68} 106005 (2003)

\bibitem{EvolutCosmologicalperturbation}
D. Langlois Phys. Rev. Lett. {\bf86} 2212 (2001)

\bibitem{Transdim phys in inflation}
G.F. Giudice, E.W. Kolb, J.Lesgourgues, Phys. Rev. D {\bf66} 083512 (2002)

\bibitem{Sub-Lunar PBH}
K. T. Inoue, T. Tanaka, Phys. Rev. Lett. {\bf91} 021101 (2003)

\bibitem{gravitoncosmologyinUED}
J.L. Feng, A. Rajaraman, F. Takayama, Phys. Rev. D {\bf68} 085018 (2003)

\bibitem{versusgraviton}
F. Leblond, Phys. Rev. D {\bf64} 045016 (2001)



\bibitem{gravitonproductioned}
K.E. Kunze, M.Sakellariadou, Phys. Rev. D {\bf66} 104005 (2002)


\bibitem{Generalized waves in any dim}
Y.N. Obukhov, Phys. Rev. D {\bf69} 024013 (2004)

\bibitem{Gravitational Radiation in D-dim} 
V.Cardoso, O.J.C. Dias, J.P.S. Lemos, Phys.Rev.D67:064026 (2003); hep-th/0212168.
\bibitem{Detecting extra dimensions with gravity wave spectroscopy}
Sanjeev S. Seahra, Chris Clarkson, Roy Maartens
gr-qc/0408032 (2004)
\bibitem{Lisa}
http://lisa.jpl.nasa.gov
\bibitem{1}
T. Kaluza, Sitzungseber. Press. Akad. Wiss. Phys. Math. Klasse {\bf 966}, (1921)
\bibitem{2}
O.Klein, Z. F. Physik {\bf37}, ~895, (1926).
\bibitem{stimadimensioni}
V.A. Kosteleck\'y, S.Samuel, Phys. Lett. B {\bf270}, ~21 (1991)
\bibitem{MKKT}
T.Appelquist, A.Chodos, P.Frund, \emph{Modern Kaluza-Klein theories}, Addison Wesley Publishing, Inc., (1987) 

\bibitem{Witten}
E. Witten, Nucl. Phys. {\bf{186}}, ~412 (1981)
\bibitem{Wesson-rep}
J.M.Overduin, P.S.Wesson  Phys. Rep. {\bf{283}}, ~303 (1997)
\bibitem{gone?} 
A. Chodos, S. Detweiler,  Phys. Rev. D{\bf 21}, 2167, (1980).
\bibitem{Sch} 
J.Scherk, J.H.Schwarz, Phys.Lett. {\bf57B}, ~463, (1975)   

\bibitem{Crem1} 
E.Cremmer, J.Scherk, Nucl. Phys. {\bf B103}, ~393, (1976)

\bibitem{Crem2} 
E.Cremmer, J.Scherk, Nucl. Phys. {\bf B108}, ~409, (1976)

\bibitem{Luciani} 
J.F. Luciani, Nucl. Phys. {\bf B135}, ~111, (1978)

\bibitem{MTW} 
C, W. Misner, K.Thorne and J. A.Wheeler, 
{\it Gravitation}, (Freeman, San Francisco), (1973)  
\end{thebibliography}
\end{document}